\documentclass[conference]{IEEEtran}
\IEEEoverridecommandlockouts

\usepackage{cite}
\usepackage{amsmath,amssymb,amsfonts}
\usepackage{algorithmic}
\usepackage{graphicx}
\usepackage{textcomp}
\usepackage{xcolor}
\def\BibTeX{{\rm B\kern-.05em{\sc i\kern-.025em b}\kern-.08em
    T\kern-.1667em\lower.7ex\hbox{E}\kern-.125emX}}

\usepackage{todonotes}
\usepackage{multirow}
\usepackage{bbding}
\usepackage[flushleft]{threeparttable}
\usepackage{makecell}
\usepackage{hyperref}
\usepackage{url}
\usepackage{subcaption}
\usepackage{xurl}

\graphicspath{{figures}}

\begin{document}

\title{RISecure-PUF: Multipurpose PUF-Driven Security Extensions with Lookaside Buffer in RISC-V}

\author{\IEEEauthorblockN{
    Chenghao Chen, 
    Xiaolin Zhang, 
    Kailun Qin, 
    Tengfei Wang, 
    Yipeng Shi, 
    Tianyi Huang, 
    Chi Zhang 
    and Dawu Gu}
\IEEEauthorblockA{\textit{Shanghai Jiao Tong University} \\
Shanghai, China \\
\{ch.chen, 
        xiaolinzhang,
        kailun.qin,
        tengfei2019,
        siponline,
        yellowskyyi,
        zcsjtu,
        dwgu\}@sjtu.edu.cn
        }
}

\maketitle

\begin{abstract}
    RISC-V's limited security features hinder its use in confidential computing and heterogeneous platforms. This paper introduces RISecure-PUF, a security extension utilizing existing Physical Unclonable Functions for key generation and secure protocol purposes. A one-way hash function is integrated to ensure provable security against modeling attacks, while a lookaside buffer accelerates batch sampling and minimizes reliance on error correction codes. Implemented on the Genesys 2 FPGA, RISecure-PUF improves at least $2.72\times$ in batch scenarios with negligible hardware overhead and a maximum performance reduction of $10.7\%$, enabled by reusing the hash function module in integrated environments such as cryptographic engines.
\end{abstract}

\begin{IEEEkeywords}
Physical Unclonable Function, RISC-V, Secure Architecture, Cryptography
\end{IEEEkeywords}

\section{Introduction}

The rapid development of the RISC-V architecture has made it an increasingly popular choice among researchers and community contributors. Its open-source nature has fostered a growing ecosystem, attracting significant attention. However, as RISC-V continues to expand, ensuring its security has become increasingly crucial.
Cryptographic engines, including Trusted Platform Modules, Secure Elements, are adopted by mainly established computer architectures such as x86 and ARM, but not well explored in RISC-V. They play a vital role in maintaining system security. These modules provide critical features like root key generation, secure boot, and hardware isolation to protect whole systems.

Physical Unclonable Functions (PUFs) are an essential component within these secure modules, serving as a foundation for protecting the root keys of computers. Companies such as Synopsys\cite{PhysicalUnclonableFunction}, NXP\cite{LPC5500SeriesArma} and Intel\cite{Intel} have integrated PUFs into their products to enhance the security. Additionally, NVidia\cite{NVidia}, Marvell\cite{Marvell} and Samsung\cite{u.sSamsungIntroducesExynos} have initiated research efforts focused on designing advanced PUFs.

Despite significant advancements in PUF technology, the integration of PUFs into RISC-V architectures still poses challenges. PUFs can be categorized into two main types: weak and strong PUFs\cite{zerroukiSurveySiliconPUFs2022}.
Weak PUFs are typically used for trusted key generation in cryptographic engines\cite{2011HOST_PUF_KG}, while strong PUFs are utilized for lightweight authentication protocols\cite{2012SP_PUF_Authentication}. In recent years, research has increasingly focused on employing strong PUFs in advanced security protocols, such as key agreement\cite{chatterjeeBuildingPUFBased2019a, qureshiPUFRAKEPUFBasedRobust2022}, authenticated encryption\cite{2023TCAD_PUF_AE} and even PKI\cite{zhangArmoredCorePKI2024b}, further expanding the potential applications of PUF technology.
With the emerging research that leverage PUFs for enhanced security, it is evident that the RISC-V architecture could benefit from a dual approach, integrating secure key generation functionalities alongside support for secure protocol implementation.

Inspired by the PUF-related Instruction Set Extensions (ISEs) in x86, such as the OASIS\cite{CCS2013_OASIS}, we evaluate and adapt its primitives to suit the RISC-V. By balancing trade-offs for hardware and reducing complexity, we design a novel ISE tailored to RISC-V’s open and modular architecture, which makes the underlying PUF mechanisms transparent to developers. By integrating multiple functionalities through PUFs, our proposed \textit{multipurpose PUF-driven security extension}, named \textit{RISecure-PUF}, addressing the growing need for improved security while maintaining compatibility with existing cryptographic engines solutions. A hash function module is incorporated for RISecure-PUF to support a variety of security functionalities while achieving cryptographically provable security. Furthermore, the randomness inherent in hash functions effectively mitigates the threat of modeling attacks, which aim to compromise PUF unclonability through statistical methods\cite{2014_DATE_PUF_Modeling_Attacks}.

Additionally, with the ongoing development of confidential computing and heterogeneous computing platforms, more PUF arrays are being utilized in new security protocols. Considering the batch sampling scenarios of PUF Challenge-Response Pairs (CRPs), we introduce a lookaside buffer to record intermediate PUF values, thereby bypassing the time-consuming error correction mechanism and achieving acceleration.

    We summarize the research contributions as follows: 
    \begin{enumerate}
        \item \textbf{A multipurpose PUF-driven security extension.} A hash function is integrated to provide multiple functionalities. Due to the random mapping of hashes, this approach first enables cryptographically \textbf{provable security} of PUFs and improves \textbf{resistance to modeling attacks}.
        \item \textbf{A~RISC-V ISE for PUFs.} This extension provides a unified CRP sample interface that is compatible with existing RISC-V cryptographic extensions, simplifying the development of secure protocols. 
        \item \textbf{A lookaside buffer for accelerating batch sampling.} The introduction of lookaside buffer significantly accelerates the batch sampling scenarios, bypassing the ECCs like Reed-Solomon and BCH.
    \end{enumerate}

    Our experiment was conducted on the Genesys 2 FPGA\footnote{\url{https://digilent.com/reference/programmable-logic/genesys-2/}}. Compared to a minimal viable PUF design, our approach requires $2.2\times$ to $6.7\times$ additional resources. While this may seem significant, practical cryptographic engines often include existing hash function modules that can be reused. As a result, the only additional resources required are negligible connection overheads for the extension.

In a single-sampling scenario, integrating the hash extension into the CRP sampling process reduces the challenge speed by $1.2\%$ to $10.6\%$. However, in batch scenarios, the use of a lookaside buffer significantly enhances efficiency. For instance, with a batch size of 16, the efficiency improves by $7.7\times$ to $11.4\times$, and this gain increases further with larger batch iterations.
In summary, our extension imposes minimal hardware resource overhead in a practical deployment, while achieving substantial efficiency gains and simultaneously enhancing both security and functionality.

\section{Background and Related Work}

\subsection{RISC-V Architecture}

The RISC-V Instruction Set Architecture is designed with modularity in mind, allowing for various extensions that enhance its capabilities based on specific requirements. One such extension is the \textit{RISC-V Cryptography Extension}\cite{cryptoeprint:2024/1323}, which enhances the cryptographic capabilities of RISC-V processors by providing efficient hardware support for widely-used cryptographic algorithms such as AES, SHA and other standard cryptographic primitives.

Due to the open-source nature of the RISC-V, it is straightforward to design new ISEs to improve system performance. In our work, PUF-related ISE are employed to enhance the security of the RISC-V.

Additionally, the use of hash functions in our design allows compatibility with the existing cryptography extension. Specifically, certain modules of the hash function can be reused, effectively reducing hardware resource consumption.

\subsection{Physical Unclonable Function}

A PUF is a hardware-specific security primitive uses the randomness found in the disorder of physical media caused by the manufacturing variation process to provide cryptographic functionalities\cite{zerroukiSurveySiliconPUFs2022}.

A PUF can be considered a black-box function, where the input is referred to as the \textit{Challenge}, and the output as the \textit{Response}. The one-to-one projection between input and output is called a \textit{Challenge-Response Pair (CRP)}. Based on the size of the CRP space that can be sampled, PUFs are categorized into strong PUFs and weak PUFs, just as previously illustrated.

\begin{figure}[h]
    \centering
    \begin{minipage}[b]{0.60\columnwidth}
      \centering
      \includegraphics[width=\textwidth]{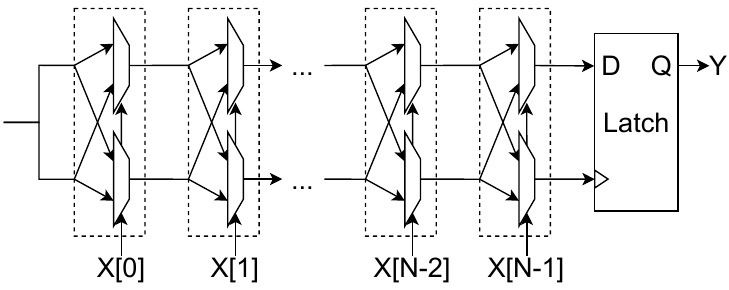}
          \subcaption{Arbiter PUF}
          \label{fig:arbiter_puf}
    \end{minipage}
    \begin{minipage}[b]{0.23\columnwidth}
      \centering
      \includegraphics[width=\textwidth]{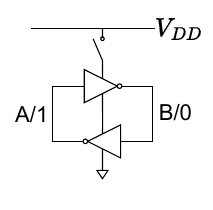}
      \subcaption{SRAM PUF}
      \label{fig:sran_puf}
    \end{minipage}
    \caption{Examples of strong PUF and weak PUF structures}
      \label{fig:puf_example}
  \end{figure}

  As shown in Fig. \ref{fig:puf_example}, the Arbiter PUF is a typical example of strong PUFs, while the SRAM PUF is a typical example of weak PUFs. The Arbiter PUF generates its response by comparing the delays of two signal paths, a process that is highly sensitive to manufacturing variations. In contrast, the SRAM PUF derives its responses from the initial states of SRAM cells, which are similarly affected by manufacturing differences. SRAM PUFs are often employed to generate secure keys, functioning as a static random entropy source. Although the both types provide physical unclonability and are practically robust\cite{PhysicalUnclonableFunction, LPC5500SeriesArma, Intel}, they lack cryptographic provability under conventional security models. Even worse, the response values of a PUF may experience bit flips.

\subsection{Fuzzy Extractor}

  To address the bit-flip issue, \textit{Fuzzy Extractor} is introduced, which could transform PUFs' noisy non-uniformly distributed secret into a stable high-entropy key\cite{10.1007/978-3-662-53140-2_20}. A fuzzy extractor consists of two phases, namely \textit{Enrollment} and \textit{Reconstruction}. In the enrollment phase, the fuzzy extractor processes the noisy PUF output to produce a stable cryptographic key. Alongside, it generates helper data, which assists in error correction but does not compromise the security of the key. In the reconstruction phase, the fuzzy extractor uses a new, slightly noisy PUF output along with the helper data to reconstruct the same stable key through Error Correction Codes (ECCs).

  When considering scenarios with multiple bit flips of PUFs, the choice of ECCs becomes critical to ensure the robustness and reliability. To handle a higher error rate, BCH code and Reed-Solomon code are often preferred due to their strong error correction capabilities.

\subsection{Controlled PUF}

The Controlled PUF (CPUF) was first introduced in \cite{1176287} with the aim of establishing a shared secret between a physical device and a remote user by utilizing control logic to manage access to the PUF. CPUF employs one-way functions to provide the PUF with multiple ``personalities", allowing the owner to regulate parameters that reveal different aspects of the PUF to various applications\cite{1176287}.

In CPUF, the hash function is primarily used for key encapsulation. In contrast, our design also employs one-way hash functions but extends the PUF's functionality to support both key generation and secure protocols, enhancing stability and resistance to modeling attacks. Unlike CPUF, our design incorporates the challenge into the subsequent hash process. Furthermore, we impose restrictions on the input and output of the hash function, ensuring improved security. In summary, while our structure is similar to CPUF, the objectives and configuration are distinctly different.

\subsection{Modeling Attacks}

Modeling attacks pose a significant threat to the security of PUF-based systems, particularly strong PUFs \cite{2014_DATE_PUF_Modeling_Attacks}. These attacks leverage machine learning techniques to create a mathematical model of a PUF's behavior by analyzing its challenge-response pairs (CRPs). Once an accurate model is constructed, an attacker can predict the PUF's responses to new challenges, effectively compromising the security of the system. To counteract these threats, non-linear features are incorporated into PUF designs, such as XOR PUF and Interpose PUF\cite{IPUF2019}, to increase complexity and resistance to modeling attacks. In our proposed RISecure-PUF architecture, the use of a one-way hash function helps mitigate the risk of modeling attacks by ensuring that the output is cryptographically secure and not easily predictable, even if an attacker gains partial knowledge of the PUF's behavior.

\section{RISecure-PUF}

\subsection{Design Overview}

\begin{figure}[htbp]
    \centering
    \includegraphics[width=1\linewidth]{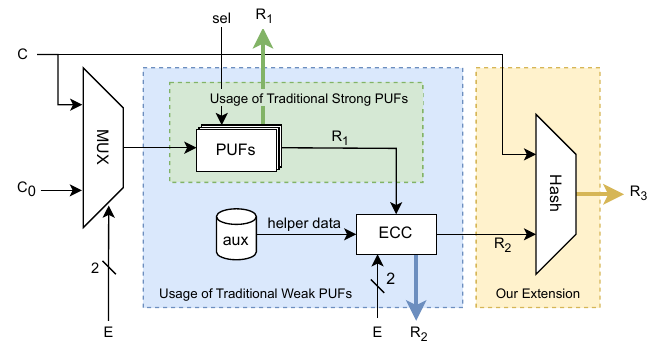}
    \caption{RISecure-PUF architecture overview}
    \label{fig:RISecure-PUF_Architecture}
\end{figure}

As illustrated in Fig. \ref{fig:RISecure-PUF_Architecture}, the proposed RISecure-PUF architecture consists of three main components: a PUF module, an Error Correction Code (ECC) module, and one-way hash function module.

In general process, the internal PUF module generates a raw response ($R_1$) based on a challenge, which is then stabilized by the ECC module (if needed) to produce a stable response ($R_2$). The stable response is used as part of the input to the hash module, where it is combined with an external challenge using direct concatenation, like $\text{Hash}(R_2 || C)$, to generate the final response ($R_3$). This design ensures that the PUF responses are stable, secure, and resistant to modeling attacks, making them suitable for cryptographic applications.

\begin{table}[htbp]
    \caption{PUF selection and output mapping in RISecure-PUF architecture}
    \label{tab:PUF_Selection_Output_Mapping}
    \begin{center}
    \begin{tabular}{|c|c|c|c|>{\raggedright\arraybackslash}m{4cm}|}
    \hline
    E{[}1:0{]} & Output & $C$           & $C_0$         & Functionality       \\ \hline\hline
    00         & R1     & \Checkmark    & \XSolidBrush  & As strong PUF for secure protocol               \\ \hline
    01         & R2     & \XSolidBrush  & \XSolidBrush  & As weak PUF for secure key generation                  \\ \hline
    10         & R3     & \Checkmark    & \Checkmark    & As strong PUF for secure protocol with higher security (modeling resilient with provable security)\\ \hline
    \end{tabular}
    \end{center}
\end{table}

The Table \ref{tab:PUF_Selection_Output_Mapping} outlines how different types of PUFs are utilized based on the selection signal \texttt{E[1:0]} and their corresponding outputs. The table specifies three configurations: 

\begin{enumerate}
    \item \texttt{E[1:0]=00}, structure directly outputs $R_1$ without requiring ECC, just like a strong PUF;
    \item \texttt{E[1:0]=01}, a weak PUF is used, generating $R_2$ for secure key generation without ECC;
    \item \texttt{E[1:0]=10}, RISecure-PUF extension operates $R_3$ via a hash function, produces the final response $R_3$. This configuration is suitable for secure protocols requiring higher security.
\end{enumerate}

Combining with Fig. \ref{fig:RISecure-PUF_Architecture}, the green section showing strong PUFs directly outputting $R_1$, the blue section demonstrating the use of PUFs with helper data and ECC to produce $R_2$, and the yellow section representing the extension module that generates $R_3$ by hashing outputs from RISecure-PUF.

In this context, the selection of the PUF module is arbitrary and can be replaced with higher-security PUFs if needed. 

The first approach involves extending strong PUFs, which inherently provide reliable outputs and therefore have a reduced dependency on ECC mechanisms. The second approach focuses on extending weak PUFs, which require greater ECC support due to their less consistent outputs.
As a result, RISecure-PUF is highly flexible, with its security dynamically adaptable to various PUF types and application scenarios.
It is only necessary to ensure that the widths of the PUF challenge and response are fixed, which will be explained in \ref{sec:prov_sec}.

From the perspective of usability, weak PUFs are more flexible and can be used in a wider range of applications. This preference is due to the simplicity and widespread adoption of typical weak PUFs, such as those based on SRAM, which are relatively easy to implement and integrate.

\subsection{Lookaside Buffer}

In security applications such as secure boot, a single PUF is often stimulated repeatedly within a short period. To improve performance in such batch scenarios, a \textit{Lookaside Buffer} is introduced to reduce reliance on ECCs like Reed-Solomon and BCH.
The primary bottleneck during batch sampling lies in the efficiency of ECCs. Due to potential bit-flip errors in each response, ECCs must repeatedly decode responses to ensure the stability of the final output. To overcome this challenge, a simple yet effective solution is to integrate a \textit{Lookaside Buffer} into the framework. Much like a \textit{Translation Lookaside Buffer (TLB)} in page table management, this buffer accelerates the process by leveraging locality in batch sampling.

\begin{figure}[h]
    \centering
    \includegraphics[width=.9\linewidth]{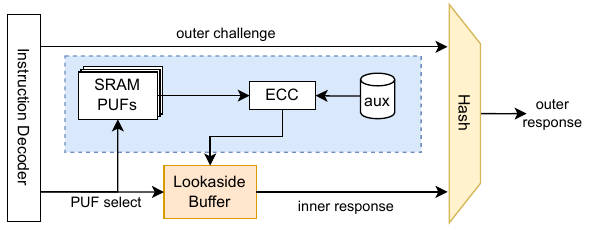}
    \caption{Lookaside buffer for batch sampling}
    \label{fig:LB_for_BS}
\end{figure}

Fig. \ref{fig:LB_for_BS} illustrates the example of SRAM-based RISecure-PUF with the lookaside buffer. The lookaside buffer stores error corrected response from inner SRAM PUFs, allowing for rapid retrieval during batch sampling. This design significantly accelerates the process, reducing the time required for ECC and enhancing the efficiency of PUF-based secure protocols.

\subsection{PUF Instruction Design}

RISC-V instruction set is designed to be modular and simple. The proposed PUF ISE follows this design philosophy, only extends necessary instructions to support PUF operations.

In \cite{CCS2013_OASIS}, the OASIS was proposed to provide an x86 platform ISE with PUFs operations. However, the OASIS is not suitable for RISC-V architecture due to its complexity and lack of modularity.

\begin{table}[h]
    \caption{OASIS PUF-related instructions and corresponding hardware primitives}
    \label{tab:OASIS}
    \begin{center}
    \begin{tabular}{|cc|cc|}
        \hline
        \multicolumn{2}{|c|}{\multirow{2}{*}{x86 Extended Instructions}}                       & \multicolumn{2}{c|}{Corresponding Hardware Primitive} \\ \cline{3-4} 
        \multicolumn{2}{|c|}{}                                                                 & \multicolumn{1}{c|}{Module}    & Micro Instruction   \\ \hline\hline
        \multicolumn{1}{|c|}{\multirow{7}{*}{OASIS}} & f\_read\_PUF                            & \multicolumn{1}{c|}{PUF}        & puf\_read           \\ \cline{2-4} 
        \multicolumn{1}{|c|}{}                       & \multirow{3}{*}{f\_init\_PUF}           & \multicolumn{1}{c|}{PUF}        & puf\_read           \\ \cline{3-4} 
        \multicolumn{1}{|c|}{}                       &                                         & \multicolumn{1}{c|}{ECC}        & ecc\_encode         \\ \cline{3-4} 
        \multicolumn{1}{|c|}{}                       &                                         & \multicolumn{1}{c|}{Crypto}     & crypto\_hash        \\ \cline{2-4} 
        \multicolumn{1}{|c|}{}                       & \multirow{3}{*}{f\_fuzzy\_extract\_PUF} & \multicolumn{1}{c|}{PUF}        & puf\_read           \\ \cline{3-4} 
        \multicolumn{1}{|c|}{}                       &                                         & \multicolumn{1}{c|}{ECC}        & ecc\_decode         \\ \cline{3-4} 
        \multicolumn{1}{|c|}{}                       &                                         & \multicolumn{1}{c|}{Crypto}     & crypto\_hash        \\ \hline
        \end{tabular}
    \end{center}
\end{table}

Table \ref{tab:OASIS} shows the OASIS instructions and corresponding hardware primitives. Since the OASIS system uses a hash module for key encapsulation, it aligns well with the extension mechanism used in our module. Therefore, a straightforward approach is to integrate the four related primitives as extended instructions, allowing for an easy implementation of the corresponding functionalities. However, we notice that excessive functional segmentation requires multiple pipeline cycles to complete a single operation, which reduces efficiency. To address this, we propose an instruction extension with 2 instructions: \texttt{inner\_puf\_init} and \texttt{outer\_puf\_chal}, which are sufficient to support the PUF operations in our architecture.

\begin{table}[htbp]

    \caption{RISecure-PUF ISE}
    \label{tab:puf-ins}

    \begin{center}
    \begin{tabular} {|ccccccccccccccccccccccccccccccccl|}

\multicolumn{7}{l}{31} &
\multicolumn{5}{l}{24} &
\multicolumn{5}{l}{19} &
\multicolumn{3}{l}{14} &
\multicolumn{4}{l}{11} &
\multicolumn{7}{r}{7} &
\multicolumn{1}{r}{0} \\
\hline

\multicolumn{7}{|c|}{funct7} &
\multicolumn{5}{c|}{rs2} &
\multicolumn{5}{c|}{rs1} &
\multicolumn{3}{c|}{funct3} &
\multicolumn{5}{c|}{rd} &
\multicolumn{7}{|c|}{opcode} & R-type Instructions \\\hline\hline

\multicolumn{7}{|c|}{0} &
\multicolumn{5}{c|}{0} &
\multicolumn{5}{c|}{rs1} &
\multicolumn{3}{c|}{001} &
\multicolumn{5}{c|}{rd} &
\multicolumn{7}{|c|}{0101011} & \texttt{inner\_puf\_init} \\

\hline

\multicolumn{7}{|c|}{0} &
\multicolumn{5}{c|}{rs2} &
\multicolumn{5}{c|}{rs1} &
\multicolumn{3}{c|}{010} &
\multicolumn{5}{c|}{rd} &
\multicolumn{7}{|c|}{0101011} & \texttt{outer\_puf\_chal} \\
\hline

\end{tabular}
    \end{center}
\end{table}

Table \ref{tab:puf-ins} shows the RISC-V ISE for PUF operations.

\textit{Instruction 1.} \texttt{inner\_puf\_init}. This instruction initializes the inner PUF module, setting up the PUF for subsequent challenges. The instruction takes the PUF's index $idx$  and inner challenge $C_0$ as the source register and the destination register as the output of auxiliary data $aux$. $R_1$ then xor-ed with the dynamic random number $r$ and encoded, to generate the $aux = \text{Encode}(R_1 \oplus r)$ for the ECC module.

\textit{Instruction 2.} \texttt{outer\_puf\_chal}. This instruction triggers the outer PUF module to generate the final response $R_3$. The instruction takes the PUF's index $idx$, outer challenge $C$ and previously stored $aux$ as the source register and the destination register as the output $R_3$. The outer challenge $C$ is then hashed concatenated with decoded $R_2 = \text{Decode}(R_1' \oplus aux)$ to produce the final outer response $R_3 = \text{Hash}(R_2 || C)$.

If a lookaside buffer is used, the values of $R_1$ and $aux$ are stored in the buffer for future use. During the data flow, the lookaside buffer is first checked for the corresponding $R_1$ and $aux$ before triggering the ECC module. The lookaside buffer is implemented as a FIFO buffer, which simplifies hardware implementation and improves efficiency by allowing quick access to recently used data.

\section{Security Analysis}

\subsection{Attack Vector}

In this paper, we mainly concern three types of adversaries as summarized in Table \ref{tab:attack_vector} with countermeasures.

\begin{table}[h]
    \caption{Attack vector of RISecure-PUF}
    \label{tab:attack_vector}
    \begin{center}
    \begin{tabular}{|ll|l|}
    \hline
    \multicolumn{2}{|l|}{Attack Vector}                & Countermeasure                             \\ \hline\hline
    \multicolumn{2}{|l|}{A1. Modeling Attacks}         & Provable Secure in \ref{sec:prov_sec} \\ \hline
    \multicolumn{2}{|l|}{A2: Cloning Attacks}          & Provable Secure in \ref{sec:prov_sec}            \\ \hline
    \multicolumn{2}{|l|}{A3: Side Channel Attacks}     &                                            \\ \hline
    \multicolumn{1}{|l|}{} & A3.1 Invasive Attacks     & Semantically Secure in \ref{sec:side_channel_sec}       \\ \hline
    \multicolumn{1}{|l|}{} & A3.2 Non-Invasive Attacks &  Not considered, but discussed in \ref{sec:side_channel_sec}                                          \\ \hline
    \end{tabular}
    \end{center}
\end{table}

\begin{table*}[ht] %
    \begin{center}
    \begin{threeparttable}
        \caption{Functionality comparison of PUFs}
        \label{tab:functionality-comparison-of-pufs}
        \begin{tabular}{|cc|cc|c|c|c|}
    \hline
    \multicolumn{2}{|c|}{\multirow{2}{*}{PUF Types}}                    & \multicolumn{2}{c|}{Functionality}                    & \multirow{2}{*}{\makecell[c]{Resistance to\\Modeling Attacks}} & \multirow{2}{*}{Reliability (\%)} & \multirow{2}{*}{Provable Security} \\ \cline{3-4}
    \multicolumn{2}{|c|}{}                                              & \multicolumn{1}{c|}{Key Generation} & Secure Protocol &                                                &                                   &                                    \\ \hline\hline
    \multicolumn{1}{|c|}{\multirow{2}{*}{Weak PUFs}}   & SRAM PUF       & \multicolumn{1}{c|}{\Checkmark}      & \XSolidBrush     & \XSolidBrush                                     & $>88$                             & \XSolidBrush\tnote{*}                        \\ \cline{2-7} 
    \multicolumn{1}{|c|}{}                             & RO PUF         & \multicolumn{1}{c|}{\Checkmark}      & \XSolidBrush     & \XSolidBrush                                    & $99.52$                           & \XSolidBrush\tnote{*}                        \\ \hline
    \multicolumn{1}{|c|}{\multirow{3}{*}{Strong PUFs}} & Arbiter PUF    & \multicolumn{1}{c|}{\XSolidBrush}    & \Checkmark       & \XSolidBrush                                    & $99.76$                           & \XSolidBrush                        \\ \cline{2-7} 
    \multicolumn{1}{|c|}{}                             & XOR PUF        & \multicolumn{1}{c|}{\XSolidBrush}    & \Checkmark       & \XSolidBrush                                    & $99.52$                           & \XSolidBrush                        \\ \cline{2-7} 
    \multicolumn{1}{|c|}{}                             & Interpose PUF\cite{IPUF2019} & \multicolumn{1}{c|}{\XSolidBrush}    & \Checkmark       & \XSolidBrush                                    & $\approx 100$                     & \XSolidBrush                        \\ \hline
    \multicolumn{2}{|c|}{RISecure-PUF}                                          & \multicolumn{1}{c|}{\Checkmark}      & \Checkmark       & \Checkmark                                      & 100                                 & \Checkmark                          \\ \hline
    \end{tabular}
    \begin{tablenotes}
        \item[*] In the literature, Weak PUFs are often used to generate secure keys and can be regarded as static random entropy sources. While their security relies on physical unclonability and is practically robust, it is not cryptographically provable under traditional security models.
    \end{tablenotes}
    \end{threeparttable}
    \end{center}
    \end{table*}

\textbf{A1: Modeling Attacks:} These attacks leverage machine learning techniques to create an accurate mathematical model of the PUF's behavior by analyzing its CRPs. Once a model is built, attackers can predict the responses to new challenges, compromising the system's security.

\textbf{A2: Cloning Attacks:}  Cloning attacks involve replicating the physical characteristics of a PUF to create a near-identical copy. By doing so, adversaries aim to produce a clone that exhibits similar challenge-response behavior, enabling unauthorized access to secure systems.

\textbf{A3: Side Channel Attacks:} Side channel attacks exploit physical information inadvertently leaked during the operation of a PUF. These attacks enable adversaries to circumvent conventional security mechanisms and gain access to sensitive information. It could be further categorized into two subtypes:

\begin{itemize}
    \item \textbf{A3.1: Invasive Attacks:} These attacks involve direct physical interaction with the PUF, often requiring specialized equipment and expertise. Examples include probing attacks, which physically access internal signals; reverse engineering, which reconstructs the PUF's structure and functionality; and fault injection, which intentionally disrupts the PUF's operation to extract information. While these methods are typically resource-intensive, they can provide detailed insights into the PUF's inner workings.
    \item \textbf{A3.2: Non-Invasive Attacks:} These attacks do not require physical modification or destruction of the PUF. Instead, they rely on analyzing indirect data, such as power consumption patterns, electromagnetic emissions, or processing delays. These methods are generally more accessible and stealthy.
\end{itemize}

\subsection{Cryptographically Provable Security}\label{sec:prov_sec}

In terms of Cryptography, Hash-based Message Authentication Code (HMAC) meet higher need for security than a simple hash function. However, in our design, a simple hash function is sufficient for our security requirements.

HMAC requires two inputs, a key ($k$) and a message ($m$), and returns an HMAC value (denoted as $t$). However, our design is considered insecure for traditional MACs because, given an HMAC construction \( \text{HMAC}(k, m) = h(k || m) \), an adversary can easily forge \( \text{HMAC}(k, m || \text{ext}) = h(k || m || \text{ext}) \) through the Merkle-Damgård construction. 

In our specific scenario, $k$ is the response from an internal PUF, and $m$ is the challenge from external. Since both $k$ and $m$ have fixed lengths, any extended message becomes invalid. As a result, there is no need for the additional complexity of an HMAC design, and a simple hash function $\text{Hash}(R_2 || C)$ is sufficient for our security requirements.

To prove the security, we need to reduce the distribution of PUF' response $\mathbf{R}$ to a truly uniformly random variable $\mathbf{U}$ using the  \textit{Hybrid Argument} technique\cite{fischlinOverviewHybridArgument2021}.

\textit{Proof Scratch.} Define initial hybrid \(\mathbf{H}_0 = \text{Hash}(R_2 || C)\), where \(R_2\) is the inner PUF's random response. Replace \(R_3\) with \(\text{Hash}(C)\) to define \(\mathbf{H}_1\), and by the Merkle-Damgård construction, \(\mathbf{H}_0\) and \(\mathbf{H}_1\) are indistinguishable. Next, define \(\mathbf{H}_2\) with \(R_3 = \mathcal{RO}(C)\), where \(\mathcal{RO}\) is a random oracle. By the random oracle assumption, \(\mathbf{H}_2\) is indistinguishable from \(\mathbf{H}_1\). Since \(C\) is random, \(\mathbf{H}_2\) and \(\mathbf{H}_0\) are computationally indistinguishable, proving the result is truly random.

Therefore, in the case of \textbf{A1}, RISecure-PUF is considered secure. If an adversary were capable of modeling RISecure-PUF, it would imply the ability to model a hash function, which contradicts random oracle model.

Further, in the case of \textbf{A2}, since the internal PUF inherently possesses unclonable properties, adversaries can only attempt to compromise the extended hash component. Additionally, by fixing the lengths of $R_2$ and $C$, it becomes impossible for adversaries to extrapolate from the hash result $R_3$, thereby ensuring that the system remains unclonable.

\subsection{Security towards Side Channel Attacks}\label{sec:side_channel_sec}

The physical unclonable nature of PUFs originates from the slight variations inherent in the manufacturing process. Therefore, any invasive attack method (\textbf{A3.1}) would alter the physical characteristics of the PUF, effectively turning it into a different instance and preventing further attacks. However, non-invasive methods (\textbf{A3.2}) may target the intermediate computational process of the PUF, potentially revealing the corresponding response values. In this paper, we do not consider countermeasures for \textbf{A3.2}.

\section{Experimental Evaluation}

\subsection{Experimental Setup and Metrics}

In this paper, a Genesys 2 Kintex-7 FPGA board (Xilinx part number XC7K325T-2FFG900C) is used to conduct evaluation experiments. Meanwhile, Xilinx Vivado 2018.2 is picked to synthesize the design with default settings.

The proposed solution was evaluated utilizing a combination of simulators and hardware experiments. Evaluation metrics include:

\begin{enumerate}
    \item \textbf{Response Reliability}: Assessing the likelihood of bit-flip errors in PUF responses. This metric is based on the data presented in \cite{zerroukiSurveySiliconPUFs2022}.
    \item \textbf{Resistance to Attacks}: Evaluating the security level achieved against various physical and modeling attacks.
    \item \textbf{Hardware Resource Consumption}: Measuring the computational resources like Look-Up Tables (LUTs) and Flip-FLops (FFs) required for response generation and stabilization.
    \item \textbf{CRP Generation Efficiency}: Evaluating the speed of CRPs generation on both single and batch scenarios.
\end{enumerate}

\subsection{Functionality Evaluation}
In Table \ref{tab:functionality-comparison-of-pufs}, we compare the functionality of different PUF types, including SRAM PUF, RO PUF, Arbiter PUF, XOR PUF, and Interpose PUF. The proposed RISecure-PUF architecture is also included in the comparison, demonstrating its superior functionality and security features in response reliability and resistance to attacks.

\subsection{Performance Overheads}

Overall, adding the hash module introduces extra hardware overhead cannot be ignored, though it has minimal impact on efficiency. However, as previously mentioned, since the hash module can be reused, this additional hardware cost can be mitigated by employing simple logic. As a result, the overall hardware overhead of the extension remains acceptable.

Table \ref{tab:resource-overheads} summarizes the hardware resource requirements for the RISecure-PUF architecture, including configurations with ECC (Reed-Solomon and BCH) and the SHA-3 module. Despite requiring approximately $2.2\times$ to $6.7\times$ more resources, much of this overhead can be offset by reusing existing hash-related modules often present in cryptographic engines.

\begin{table}[h]
    \caption{Hardware resource overheads of RISecure-PUF}
    \label{tab:resource-overheads}
    \begin{center}
        \begin{tabular}{|l|l|l|}
            \hline
            PUF Types     & \#LUTs & \#FFs \\ \hline\hline
            SRAM PUF-RS     & $1,854$  & $2,157$ \\ \hline
            SRAM PUF-BCH             & $634  $  & $ 509$ \\ \hline
            RISecure-PUF-RS                       & $5,457 (2.94\times)$  & $4,832 (2.24\times)$ \\ \hline
            RISecure-PUF-BCH                     & $4,237 (6.68\times)$  & $3,184 (6.26\times)$  \\ \hline
            \end{tabular}
    \end{center}
\end{table}

Efficiency comparisons in Table \ref{tab:SRAM-RISC-comparison} show that RISecure-PUF brings only a minor performance reduction, with a decrease of $1.2\%$ to $10.6\%$, depending on the ECC configuration. This demonstrates that the additional security features have a negligible impact on performance.

\begin{table}[h]
    \caption{Efficiency comparison of SRAM PUF and RISecure-PUF}
    \label{tab:SRAM-RISC-comparison}
    \begin{center}
        \begin{tabular}{|c|cc|}
            \hline
            \multirow{2}{*}{PUF types} & \multicolumn{2}{c|}{CRPs sampled per millisecond (ms$^{-1}$)}\\ \cline{2-3} 
                                       & \multicolumn{1}{c|}{Reed-Solomon} & BCH   \\ \hline\hline
            SRAM PUF                   & \multicolumn{1}{c|}{$66.55$}      & $173.4$\\ \hline
            RISecure-PUF                       & \multicolumn{1}{c|}{$59.46$}      & $171.3$\\ \hline
            Comparison                 & \multicolumn{1}{c|}{$-10.66\%$}     & $-1.16\%$\\ \hline
            \end{tabular}
    \end{center}
\end{table}

\begin{figure}[h]
    \centering
    \includegraphics[width=.9\linewidth]{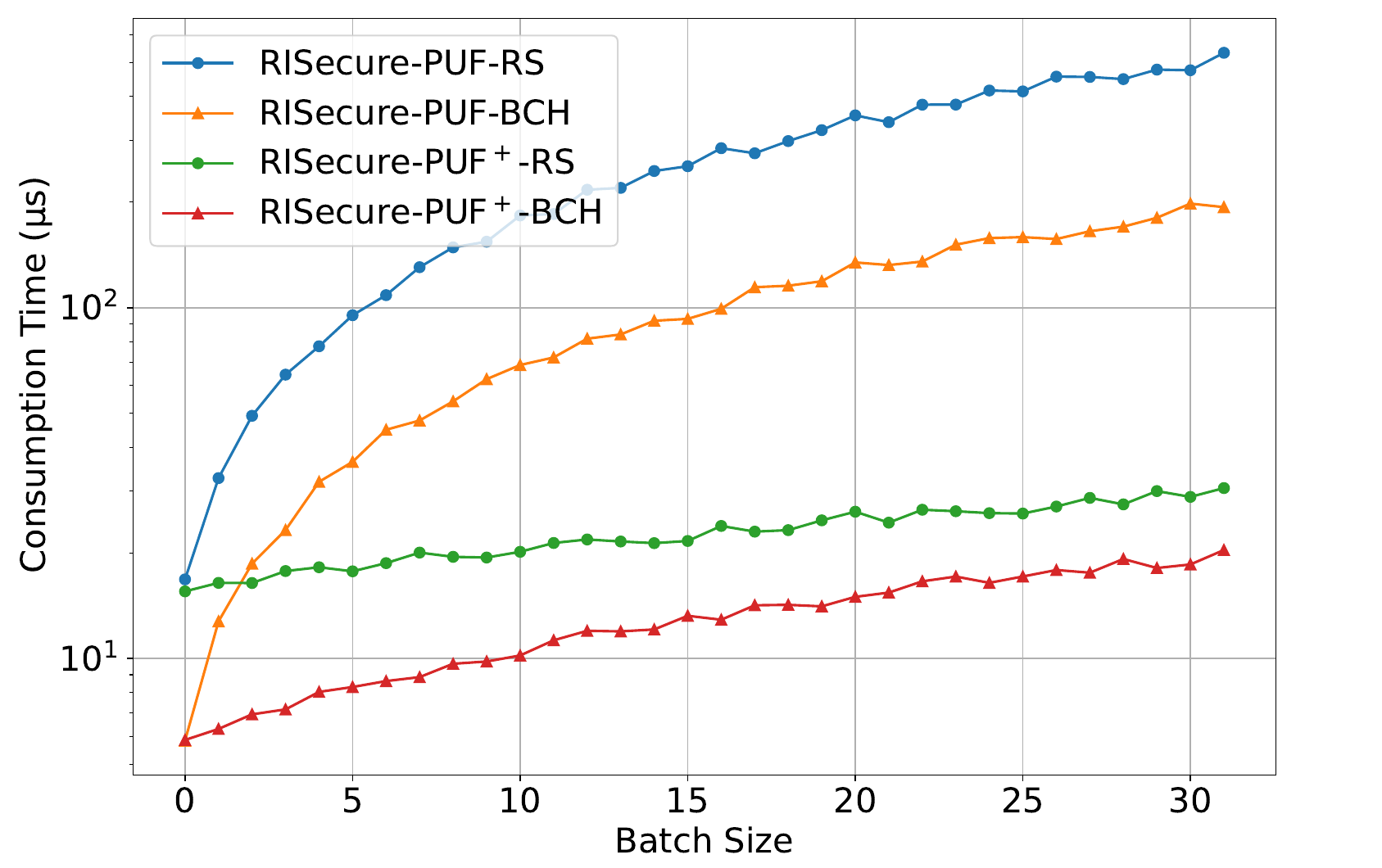}
    \caption{Performance evaluation of extra lookaside buffer}
    \label{fig:performance}
\end{figure}

To further enhance performance, a lookaside buffer was introduced to optimize the batch sampling process. As shown in Fig. \ref{fig:performance}, the buffer (refer to RISecure-PUF$^+$) significantly accelerates sampling time of batch scenarios.

We can clearly observe that, in the logarithmic scale, the growth of RISecure-PUF$^+$ is nearly linear. This indicates that as the batch size increases, the time savings achieved by the improvement may eventually surpass the total time consumed by the original design. However, considering that such large batch sizes are unrealistic in practical scenarios, we use the case with a batch size of 16 as a fair and representative example.

Table \ref{tab:RISC-RISC-p-comparison} gives an example with a batch size of 16. The lookaside buffer could speed up the sampling by $2.72\times$ for Reed-Solomon and $1.63\times$ for BCH, which represents a significant improvement.

\begin{table}[h]
    \caption{Time comparison of original RISecure-PUF and RISecure-PUF$^\text{+}$ with lookasdie buffer}
    \label{tab:RISC-RISC-p-comparison}
    \begin{center}
        \begin{tabular}{|c|cc|}
            \hline
            \multirow{2}{*}{PUF types} & \multicolumn{2}{c|}{16 Rounds Time Consumption ($\mu$s)}\\ \cline{2-3} 
                                       & \multicolumn{1}{c|}{Reed-Solomon} & BCH   \\ \hline\hline
            RISecure-PUF                   & \multicolumn{1}{c|}{$253.13$}      & $21.64$\\ \hline
            RISecure-PUF$^\text{+}$        & \multicolumn{1}{c|}{$92.94$}      & $13.24$\\ \hline
            \end{tabular}
    \end{center}
\end{table}

In conclusion, while the RISecure-PUF introduces modest hardware and performance overheads, these are well within acceptable limits, especially with optimizations like the lookaside buffer, which remarkably enhance its practicality for secure and resource-constrained applications.

\section{Conclusion}

This paper presents RISecure-PUF, a security extension for RISC-V that integrates PUFs to provide multiple secure functionalities (key generation and secure protocol) and enhance the reliability towards modeling attacks. By combining existing PUFs with a one-way hash function, the design achieves provable security against modeling attacks while maintaining compatibility. The inclusion of a lookaside buffer significantly optimizes batch CRP sampling, through bypassing time-consuming ECCs. Our analysis and experimental evaluations demonstrate improved reliability, security, and efficiency, making RISecure-PUF a practical and scalable solution for secure, resource-constrained environments. Future work could extend this framework to other architectures and explore new optimizations for broader applicability.

\newpage
\bibliographystyle{IEEEtran}
\bibliography{IEEEabrv,main}

\end{document}